\documentclass[
showpacs,
floatfix,
aps,prb,amsmath,
twocolumn,
groupaddress,
]{revtex4}

\usepackage{epsfig}
\usepackage{bm}
\usepackage{epsf}
\usepackage{array}

\newcommand{\be}{\begin{equation}}
\newcommand{\ee}{\end{equation}}

\newcommand{\bea}{\begin{eqnarray}}
\newcommand{\eea}{\end{eqnarray}}

\newcommand{\bes}{\begin{split}}
\newcommand{\ees}{\end{split}}

\newcommand{\req}[1]{Eq.~(\ref{#1})}

\newcommand{\Ec}{E_{\rm c}}
\newcommand{\kB}{k_{\rm B}}
\newcommand{\varef}{E_{\rm F}}
\newcommand{\vare}{\varepsilon}
\newcommand{\mls}{\delta_1}
\newcommand{\Nl}{N_{\rm l}}
\newcommand{\Nr}{N_{\rm r}}
\newcommand{\Nch}{N_{\rm ch}}

\begin{document}
\title{Failure of the Wiedemann--Franz Law in Mesoscopic Conductors}
\author{Maxim G. Vavilov and A. Douglas Stone}
\affiliation{Department of Applied Physics, Yale University, New
Haven, CT 06520}

\date{\today}

\begin{abstract}
We study the effect of mesoscopic fluctuations on the validity of
the Wiedemann--Franz (WF) law for quasi-one-dimensional metal wires
and open quantum dots.
At temperatures much less than the generalized Thouless energy,
$E_c$, the WF law is
satisfied for each specific sample but as the
temperature is raised through $E_c$  a sample-specific correction to
the WF law of order $1/g$ appears ($g$ is the dimensionless conductance) and
then tends to zero again at $k_BT \gg E_c$. The mesoscopic violation
of the Weidemann-Franz law is even
more pronounced in a ring geometry, for which  Lorenz number
exhibits $h/e$ flux-periodic Aharonov-Bohm oscillations.
\end{abstract}


\maketitle

\section{Introduction}

The Wiedemann-Franz (WF) law \cite{WF} in macroscopic conductors relates the
electrical and thermal conductivities $\sigma,\kappa$ via the
condition $\kappa/\sigma T= l_0$, where the Lorenz number $l_0
\equiv (\pi^2/3) (\kB /e)^2$ (here $\kB ,e$ are Boltzmann's constant
and the electron charge).  The law is satisfied with an accuracy
of $5-10\%$ in most metals at room temperature~\cite{Kittel},
but fails in the temperature range $T \approx 10 - 100$K where
%%%^{\circ}
the Bloch-Gruneisen form of the resistivity holds  \cite{madelung-214}.
In this range the failure is understood because inelastic
phonon scattering degrades thermal and electrical currents
differently. At temperatures below $10$K, when the %%%phonon
inelastic
contribution to the scattering is negligible the law is satisfied
extremely accurately \cite{uher78} as it follows simply from the
Sommerfeld expansion of the energy-dependent conductivity, which
is accurate to order $(\kB  T/\varef)^2$, where $\varef$ is the Fermi
energy.  Failure of the WF law at low temperatures is often
associated with non-Fermi-liquid behavior in
strongly correlated systems~\cite{nature414}.

In the current work we focus on the violation of the WF law induced
by mesoscopic fluctuations in Fermi-liquid systems.
Mesoscopic conductors are normal conductors on the micron or nanometer scale
connecting bulk leads which act as thermal equilibrium
reservoirs held at different voltages and
temperatures. By assumption there are no inelastic scattering
processes in the conductor itself, exactly the condition under which the
WF law holds for macroscopic conductors. However, as is well
known, each specific mesoscopic sample exhibits a fluctuating
energy-dependence of its conductance which varies by $\delta G \sim
e^2/h$ on an energy scale $ \Ec  \ll \varef$ \cite{UCF,UCF_2}.
Therefore the Sommerfeld expansion fails even when $\kB T \ll \varef$.
The most immediate effect of this failure is that the thermopower
coefficient, which gives the voltage induced by a temperature
gradient when no electrical current flows, becomes random in sign
and is enhanced by $(\varef/ \Ec )g^{-1}$, where $g \equiv
G/G_0$ is the dimensionless conductance and $G_0=e^2/h$.
This enhancement can be many orders of magnitude, has been discussed
in a number of
previous works \cite{AAAZ,LK88,esposito,LSB,moskal}, and has been
measured in at least one experiment on quantum dots \cite{godijn}.

It is straightforward to show that the
violation of the WF law is
{\it not} due to the enhanced thermopower coefficient and
occurs at
size scales for which the thermopower effect would be negligible.
As usual, the mesoscopic correction to the WF law is random and arises
from interference of multiply-scattered electron trajectories; in a
ring geometry this correction will oscillate periodically in magnetic flux with
period $h/e$, making it easier to detect.  Very recently
thermoelectric measurements have been made in such systems
\cite{chandrasekhar},
so experiments of this type are feasible.

We consider a two-probe measurement for which
the electrical current, $I_{\rm e}$, and thermal current,
$I_{\rm h}$, between the reservoirs at temperatures $T$
and $T+\delta T$ and at voltage  $V$
are
\begin{subequations}
\label{matrix}
\bea
I_{\rm e} & = & G V + B\delta T ,\\
I_{\rm h} & = &  \Gamma V +\Xi \delta T.
\label{Iq}
\eea
\end{subequations}
The off-diagonal coefficients, $B$ and   $\Gamma$,
are related by Onsager's principle: $\Gamma =-TB$.

To the leading order in $(E_{\rm F}\tau)^{-1}$ the electron--electron
interaction does not affect the fluctuations of transport
coefficients at low temperature ($\tau$ is the mean
elastic scattering time)~\cite{LSF87}; in addition inelastic scattering
due to interactions at finite temperature may be neglected in the
mesoscopic regime. Interactions in disordered metals do affect
the average electric and thermal
conductances and can lead to a violation of the WF
law~\cite{Smith,BAA05}; however this effect does not have magnetic
flux sensitivity and its temperature dependence is different from
that of the mesoscopic violation studied here.

Neglecting the interaction effects, the thermoelectric coefficients can all be
expressed in terms of integrals over $G(\vare)$, where the $T=0$
conductance
at energy $\vare$ is given by the Landauer-B\"uttiker
formula
$G(\vare) =2G_0 \sum_{\alpha,\beta} | t_{\alpha \beta}(\vare)|^2$
and
$t_{\alpha \beta} (\vare)$ is the transmission amplitude for
conduction channels
$\alpha$, $\beta$.
Defining the energy moments of the conductance $G(\vare)$ at
temperature
$T$ and chemical potential $\mu$ by
\be
G^{(n)} (T) = -\int_{\infty}^{\infty}\!\!\!\! J_n(\vare) d\vare
,
\  J_n(\vare)\equiv
G(\vare)
\frac{(\vare - \mu)^n}{(\kB T)^n}
\frac{df}{d\vare}
\label{Gn}
\ee
one finds that $G = G^{(0)}$, $B=-(\kB /e) G^{(1)}$ and
$\Xi = (\kB /e)^2 T G^{(2)}$.
   The thermal  conductance $G_{\rm h} \equiv I_{\rm h}/\delta T$,
measured when $I_{\rm e}=0$.
  From \req{matrix} we have
$G_{\rm h} = \Xi - \Gamma B/G$.
The WF law has the same form for these conductance coefficients as
for the
bulk conductivities: $G_{\rm h}/GT =l_0$.

If $G(\vare)$ is constant on scales much greater than $k_BT$ then
the thermal conductance $G_{\rm h}$ obeys the WF law to very high
accuracy.  First, if
$G(\vare)$ is constant and hence equal
for electrons and holes within $\kB  T$ of the Fermi energy then
no net electrical current flows between the two reservoirs under a
thermal gradient and the
thermopower is zero: $\Gamma = -TB= 0$.
Thermal current {\it does} flow
since the colder reservoir receives more electrons with energy
$\vare - \mu >0$ and holes with energy $\vare - \mu <0$;
this flux is given by setting $G(\vare) = G(\mu)\equiv G_\mu$, leading immediately to
$G^{(2)} = (\pi^2/3) G_\mu $ and $G_{\rm h}/G=\Xi/G = l_0T$.  Thus the
WF law does hold as $T \to 0$ for mesoscopic conductors;
fluctuations in both thermal and electrical
conductances maintain a fixed ratio given by the WF law.

\begin{figure}
\epsfxsize=0.37\textwidth
%\vspace*{0.3\textwidth}
\centerline{\epsfbox{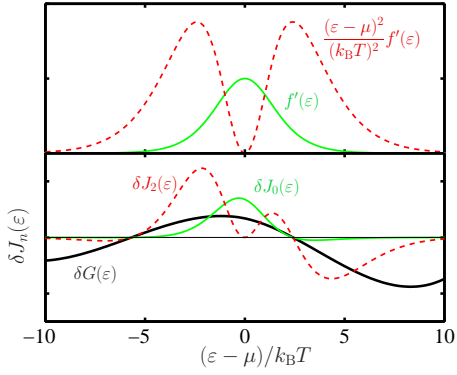}}
\caption{
Upper panel shows the thermal weight functions for the electrical
current, $I_{\rm e}$,
and for the heat current, $I_{\rm h}$. For
constant electron transmissivity
$  \sim G (\vare)$ on the scale of
$k_BT$
the ratio of areas below these functions leads to the WF law, which thus
always holds as $T \to 0$.
However if $\delta G (\vare)$ fluctuates
on the scale of
$\Ec\sim k_BT$ (lower panel) then  $\delta G(\vare)$ is unequally
weighted for $I_{\rm e}$ and $I_{\rm h}$
and sample-specific corrections to the WF law appear.
}
\label{fig:1}
\end{figure}

If $G(\vare)$ does vary on the scale of $k_BT$ then one
cannot
approximate $G(\vare)$ by $G_\mu$, the thermal and electrical
currents have
different weighting factors (see Fig. 1), and their
ratio is no longer fixed at the WF value.
Thus mesoscopic conductors
will begin to violate the WF  law at temperatures which
are on the
scale of variation of $G(\vare)$.  This scale is
$ \Ec   = h/\tau_{\rm t}$, where $\tau_{\rm t}$ is the typical
transit time between the reservoirs
and depends on the geometry of the conductor~\cite{UCF,UCF_2}.
Since the mesoscopic
conductance fluctuates by $\delta G \sim e^2/h$ on this
scale we can
immediately estimate the maximum size of the violation
of the WF law.  Define
the Lorenz ratio to be
\be
l = \frac{G_{\rm h}}{GT}  = \frac{\Xi}{GT} - \frac{\Gamma B}{TG^2},
\label{lor}
\ee
where $G= G_{\mu} + \delta G(T)$ and $\Xi = l_0 T G_{\mu} +
\delta \Xi (T)$.  The
fluctuating contributions $\delta G (T), \delta
\Xi (T)$ are $ \sim g^{-1}$; therefore the fluctuating part
of the
first term in  \req{lor} is $\delta (\Xi/G)  = l_0 [\delta \Xi / l_0
G_{\mu} T  - \delta G/G_{\mu}]
\sim g^{-1}$.   In contrast, the
thermopower coefficients  $\Gamma , B$ are zero in leading order
and
only have a contribution from the mesoscopic fluctuations
\cite{AAAZ,LK88,LSB};
therefore the fluctuating part of the
second
term in \req{lor} is $\Gamma B/G^2 \sim g^{-2}$ and the thermopower
fluctuations do not
contribute to the violation of the WF law at
leading order.  Thus the fluctuations $l_0\eta$ in
the Lorenz ratio $l$ are given
by
\be
    l=l_0(1+\eta)  =   \frac{\Xi}{GT}  ,  \;\;\; \eta =
\frac{1}{G_\mu }\left[\frac{\delta \Xi}{l_0 T} - \delta G\right].
\label{Lorentz}
\ee

The standard deviation of the dimensionless quantity $\eta$ gives
us the typical
violation of the WF law.  Before averaging
for $\kB
T \ll  \Ec $ we can expand $G(\vare)$ in \req{Gn} around
$\mu=\varef$ to order $(\vare - \mu)^2$; the zeroth order yields the
WF law, the first order vanishes by symmetry, and the second order
will give a correction $\eta \sim (\kB  T)^2 G^{\prime
\prime}(\mu)/G_\mu$, which for a mesoscopic sample will be
$  \sim
(\kB T/ \Ec )^2 g^{-1}$.   Hence the deviation from the WF law will
rise quadratically
for $k_BT \ll E_c$ and become $\sim g^{-1}$ at
$k_BT \sim E_c$.  When
$\kB T \gg  \Ec $ the typical violation will again decrease as the
random contributions from energies $\vare$
begin to self-average over the window $k_BT$.  The
quantitative behavior of the
fluctuations can be obtained from the
variance of $\eta$, which is determined by
the zero temperature
conductance correlation functions
$K((\vare - \vare')/\Ec) = \langle\delta G(\vare ) \delta
G(\vare') \rangle/G_0^2$ as
%%From \reqs{Gn} and
%%\rref{Lorentz} we see that
\be
{\rm var}(\eta)=  \frac{G_0^2}{\langle G_\mu\rangle^2}
\left[
\frac{9C_{22}}{\pi^4}-  \frac{6C_{20}}{\pi^2  } +
C_{00}\right] \equiv   \frac{f^2(T, \Ec )}{g^2},
\label{varetaG}
\ee
where the coefficients $C_{nm}=\langle \delta G^{(n)} \delta
G^{(m)}\rangle /G_0^2$
refer to integrals of the form:
\be
C_{nm}=\!\!\! \int\limits_{-\infty}^{\infty}\!\! d \vare d\vare'
\frac{df}{d\vare}
\frac{df}{d\vare'}
\frac{(\vare - \mu)^{ n}(\vare' - \mu)^{ m}}{(\kB T)^{n+m}}
K\!\! \left[\frac{\vare - \vare'}{\Ec}\right].
\label{Cdef}
\ee
As already noted, the typical value of the violation of
the WF law is
${\rm  rms} (\eta) = f(T,  \Ec )/g$; the function $f(T,E_c)$
is plotted for the geometry of a
quasi-one-dimensional metallic wire  and an open quantum dot
in Fig. 2.  As expected, the maximum of $f(T, \Ec )$ occurs
at $\kB  T\sim  \Ec $ and $f(T, \Ec )$ vanishes for both
$\kB T \ll  \Ec $ and for $\kB  T \gg  \Ec $.
We now provide a more detailed analysis of the results.

\section{Thermal conductance}

\subsection{Quasi-one-dimensional diffusive conductors}

For a diffusive mesoscopic system (i.e., a system with the elastic
mean free path shorter than
all sample dimensions)  the conductance
correlation function $K(\Delta \vare/ \Ec )$ has been calculated
by impurity-averaged perturbation theory  \cite{UCF,UCF_2}
and can be expressed in terms of sums over the eigenvalues of the
diffuson and cooperon propagators \cite{obzor,stone91}.
For the case of a disordered quasi-one-dimensional wire of length $L$
with diffusion coefficient $D$, we have
$\Ec=D/L^2$ and
\be
K(x)=\sum_{n=1}^{\infty} \frac{16 {\cal K}(n^2,x) }{\beta \pi^4};
\ {\cal K}(\xi,x)=
\frac{3}{\xi^2+x^2}-\frac{2x^2}{(\xi^2+x^2)^2},
\label{Kx}
\ee
where the parameter $\beta = 1$ in the presence of time-reversal
symmetry and $\beta = 2$
in  its absence, and we have included spin degeneracy.
Time-reversal symmetry is effectively broken by a magnetic field comparable to
a flux $\Phi_0=h/e$ through the wire, typically rather weak fields.
Experiments which rely
on varying magnetic field to observe fluctuations will
correspond to the case $\beta =2$.
The function $f(T, \Ec )$ giving ${\rm rms}(\eta )$ for a wire is
plotted as the solid line
in Fig.~2; it has the asymptotes:
\begin{subequations}
\label{asymptotes1d}
\begin{eqnarray}
\!\!\!\! f(T, \Ec )  & = & \frac{64}{5}\sqrt{\frac{7\zeta(12)}{3}}
\left[\frac{\kB T}{ \Ec }\right]^2; \quad
\kB T\ll \Ec,
\\
\!\!\!\! f(T, \Ec )  & = &  \frac{4}{3\pi} \sqrt{\frac{4\pi^2-30}{5\pi}}
\sqrt{\frac{\Ec }{\kB T}}; \quad
\kB T\gg \Ec .
\label{f1dHT}
\end{eqnarray}
\end{subequations}
The maximum  of $f(T, \Ec )$ is $f^{\rm max}(T, \Ec )\approx 0.23$
at $\kB T_{\rm max}\approx 0.5 \Ec $ leading to a violation of the WF
law
$\sim 0.23/g$.  This expression describes both
micron scale and
nano scale metallic wires as well as semiconducting quantum wires
in
the diffusive regime for $g \gg 1$.   For such systems the
dimensionless conductance $g$ can vary
between $10^0 - 10^3$.
When $g$ approaches unity it will be necessary to include the
thermopower corrections to the WF law as well and the violation,
while order unity, will not
be quantitatively described by
$f(T,E_c)$.

\begin{figure}
\epsfxsize=0.34\textwidth
%\vspace*{0.3\textwidth}
\centerline{\epsfbox{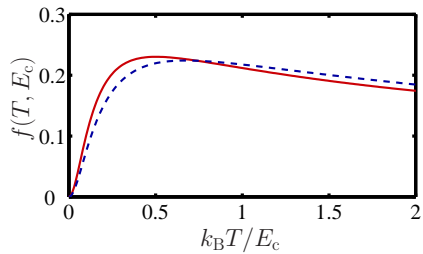}}
\caption{ Temperature dependence of the $\rm{rms}(\eta)$ in
quasi-one-dimensional wires (solid line) and in open quantum dots
with $\Nl=\Nr\gg 1$ (dashed line).  In both cases a broad maximum
in
the violation of the WF law occurs at $k_BT \sim E_c$ and persists
to
much higher temperatures.
}
\label{fig:2}
\end{figure}

\subsection{Aharonov--Bohm oscillations }

In order to make the mesoscopic deviation from the WF law
more easily measurable it will be convenient to fabricate
ring structures
similar to those used to first measure the mesoscopic $h/e$-periodic
Aharonov-Bohm (AB) effect in normal metals \cite{UCF_exp,AS}.
In this case the mesoscopic quantum corrections to the conductance
are the only terms which give rise to oscillations of the
conductance with period $h/e$ and these terms are relatively easy to
isolate by Fourier transforming the measured behavior of $G(\Phi)$
and $G_{\rm h} (\Phi) \approx \Xi (\Phi)$, where $\Phi\propto B$ is the
magnetic flux through the ring in magnetic field $B$.
We consider then the ratio of harmonics
of the AB oscillations at frequency $k/\Phi_0, k = 1,2,\ldots$ in
$\delta G(\Phi)$ and $\delta \Xi(\Phi)$.
For $\kB T \ll  \Ec $ we have
the simple result that
$\delta G(\Phi) \approx \delta G_\mu(\Phi)$ and
$\delta \Xi \approx l_0T \delta G_\mu (\Phi)$;
hence $\delta \Xi (\Phi)/ \delta G(\Phi) = l_0 T$ for all
$\Phi$.   The AB oscillations are perfectly in phase and their
ratio is exactly the Lorenz number (as is the ratio of each
harmonic).  However for $\kB T \sim  \Ec $ the oscillating
contributions to $G$, $\Xi$ are no longer proportional to one another
and their ratio need not be the Lorenz number.  To quantify their
relationship we calculate the harmonics of the conductance and
heat conductance correlation functions (in flux) for such a ring.
Let  ${F}_{G}(\Delta\Phi)= \langle \delta G(\Phi) \delta
G(\Phi+\Delta\Phi)\rangle$ and ${F}_{\Xi}(\Delta\Phi)=
\langle \delta \Xi(\Phi) \delta  \Xi(\Phi+\Delta\Phi)
\rangle$; neglecting the flux through the wires making up the ring
(i.e. including only the flux through the hole) these functions
are periodic with fundamental frequency $\Phi_0$
and can be represented in terms of their  harmonics $F^{(k)}_{G,\Xi}$
\be
{F}_{G,\Xi}(\Delta\Phi)=\alpha
\sum\nolimits_{k=0}^{\infty}
F^{(k)}_{G,\Xi}\cos\left(\frac{2\pi k \Delta\Phi }{\Phi_0}\right),
\label{etacorr}
\ee
where $\alpha$ is geometry dependent factor. As above, \req{Kx}, the
relevant correlation functions can be expressed as sums where for
the ring the integers in the sum appear only in the combination $m
- \Delta \Phi/\Phi_0$ (leading to an obvious periodicity in
$\Phi_0$).  It is convenient in this case to rewrite the sum using
the Poisson summation formula where the integration variable is
shifted to $\phi = m - \Delta  \Phi /\Phi_0$ \cite{AS}.  This yields a
relatively simple formula for
for the harmonics
$F_{G}^{(k)}= G_0^2 F^{(k)}_{00} $ of the electrical conductance
and for the harmonics
$F_{\Xi}^{(k)}= (\kB^2T/h)^2 F^{(k)}_{22} $
of the thermal conductance. Here
\be
\begin{split}
F^{(k)}_{nm}= &
\int\nolimits_{\infty}^{\infty}  d\vare d\vare'
\frac{df}{d\vare} \frac{df}{d\vare'}
\frac{(\vare  - \mu)^n (\vare' - \mu)^m}{(\kB T)^{n+m}}
\\
& \times
\int\nolimits_{-\infty}^{\infty} d\phi \cos(2\pi k \phi)
{\cal K} \left(\phi^2+\frac{\gamma}{E_{R}},\frac{\vare
-\vare'}{E_R}\right)
\label{harmonicsX}
\end{split}
\ee
with kernel ${\cal K}(\xi,x)$ defined by \req{Kx},
$  E_{R}=D/R^2$ is inversely proportional to the electron diffusion
time along a circumference of the ring of radius $R$, and $\gamma$
takes into account electron escape from the ring through the leads
and other phase relaxation rates, such as electron-electron or
electron-phonon dephasing rates.

\begin{figure}
\epsfxsize=0.38\textwidth
%\vspace*{0.3\textwidth}
\centerline{\epsfbox{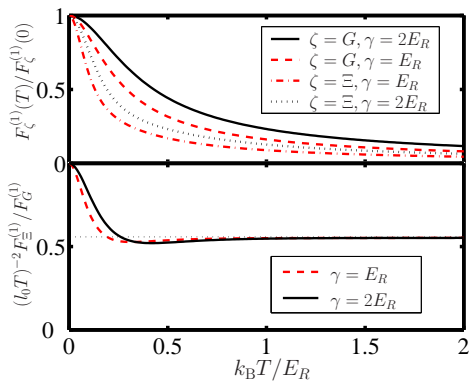}}
\caption{
(Top) Temperature dependence of $F^{(k=1)}_{G,\Xi}(T)$ for
two values of $\gamma /E_{R}$. At high temperatures
$\kB T\gg E_{ R}$ amplitudes $F^{(k=1)}_{G,\Xi}(T)$ decrease only
as $T^{-1}$. (Bottom) the
ratio $F^{(1)}_\Xi/F^{(1)}_G$ as a function of temperature.
At
$T\gtrsim E_{R}$ the ratio saturates to  $0.55 (l_0 T)^2$
(dotted line), see \req{Lnew}.
}
\label{fig:3}
\end{figure}

Performing the integrations in \req{harmonicsX},
we find that the ratio of the harmonics of
$\Xi$ to $G$ decreases from the
WF value monotonically with increasing temperatures until it saturates at
$\kB T\gtrsim    E_{R},\   \gamma$, to the value
\be
\frac{1}{T^2}
\frac{F_{\Xi}^{(k)}}{F_{G}^{(k)}}=\frac{21\pi^2-180}{5\pi^2}
l_0^2 \approx 0.55 l_0^2.
\label{Lnew}
\ee
In this temperature regime the conductance and heat conductance
AB oscillations will appear uncorrelated and their ratio will randomly
vary at each value of magnetic flux $\Phi$; it is only the ratio
of their variances (or harmonics)
averaged over magnetic fields which
will saturate to a constant value.  This saturation value will be
different by roughly
a factor of one half from the value expected from the WF law.
The reason that the saturation value is less than the WF value when
$\kB T > \Ec $ is
that the conductance fluctuations are determined by transmission
fluctuations within
$\kB T$ of the Fermi level, while the thermal conductance
fluctuations are determined
by transmission fluctuations in two regions of width $\kB T$
displaced from the Fermi level by $\sim \pm \kB T$,  see Fig. 1.
Hence the contributions from different energies to
the thermal conductance fluctuations are less correlated and tend
to cancel each other more
than the contributions to the electrical conductance fluctuations.
We plot the ratio $F_{\Xi}^{(k)}/F_{G}^{(k)}$ as a function
$\kB T/  E_{R}$ for several values of $ \gamma$ in Fig. \ref{fig:3}.

\subsection{Open Quantum Dots }

If we consider semiconducting systems the above calculations also apply and
since $g$ is typically much smaller one expects large violations of
the WF law, although
in general the WF law is not well-satisfied in semiconductors  so the
novelty is less than
in metals.   Semiconducting quantum dots exhibit large
aperiodic thermopower fluctuations as a function of magnetic
field\cite{godijn}; these
have been treated theoretically by Van Langen et  al. \cite{LSB}.   The
behavior of the Lorenz number for this case was not studied either
experimentally
or theoretically.  As long as the number of open
channels $N \gg 1$ our \req{varetaG} for the variance of $\eta$
still applies with the only change
being the replacement of \req{Kx} for the conductance energy
correlation function by the
appropriate function for an open quantum dot.  If we assume the quantum dot
junction generates chaotic scattering then the S-matrix of the system will be
described approximately by the Dyson ensembles of random matrix theory
and the energy correlation function
$
K _{\rm qd}(\Delta\vare/\Ec)= (\Nl^2\Nr^2/\Nch^4) (1+\Delta\vare^2/\Ec^2)^{-1}
$,
where $ \Ec   = \Nch\mls/(2\pi)$,
$\Nl,\Nr$ are the number of incoming channels in the left and right
leads of the
junction, $\Nch=\Nr+\Nl$, and $\mls$ is the mean level spacing of electron
states in the dot~\cite{beenakkerRMP}.
Substituting $K _{\rm qd}(x)$ into \req{Cdef},
we calculate ${\rm var }(\eta)=f_{\rm qd}^2(T,\Ec)/g^2$
with $g=2\Nl\Nr/\Nch$ from \req{varetaG}.
$f_{\rm qd}(T,\Ec)$ has the following asymptotic forms:
\begin{subequations}
\begin{eqnarray}
\!\!\!\!\! \!\!\!\!\!
f_{\rm qd} & = & \frac{16 \pi^2}{5} \sqrt{\frac 23} \frac{\Nl\Nr}{\Nch^2}
\left[\frac{\kB T}{ \Ec }\right]^2 ,\quad \kB  T \ll \Ec ,
\label{etaTlow}
\\
\!\!\!\!\! \!\!\!\!\!
f_{\rm qd} & = & 2\sqrt{\frac{2\pi^2-15}{15 \pi}} \frac{\Nl \Nr }{\Nch^2}
\sqrt{\frac{ \Ec }{\kB T}},\quad \kB T \gg  \Ec ,
\label{etaThigh}
\end{eqnarray}
\end{subequations}
and reaches its maximum value $f_{\rm qd}^{\rm max}\approx 0.88 \Nl\Nr/\Nch^2$
at $\kB T_{\rm max}\approx 0.68  \Ec $.  We plot the function $f_{\rm
qd}(T,\Ec )$
for a chaotic quantum dot (dashed line) in comparison
to the wire case in Fig. 2; the behavior of the two cases is quite similar.
Note that in both cases
the decay at $\kB  T \gg \Ec$ is very slow $\sim T^{-1/2}$ and the maximum is
very broad
on the high temperature side.  Finally we note that for a non-chaotic quantum
dot or metallic nanobridge \cite{nbridge} the conductance
fluctuations can exceed $G_0$ and consequently the violations of
the WF law can be even larger
than we find here.  The behavior for specific non-chaotic shapes
is non-universal and can be
calculated by semi-classical methods \cite{BJS}.

\begin{figure}
\epsfxsize=0.34\textwidth
%\vspace*{0.3\textwidth}
\centerline{\epsfbox{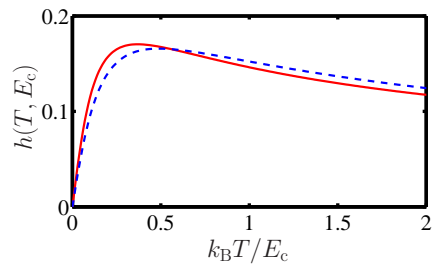}}
\caption{Temperature dependence of the $\rm{rms}(B)$ in
quasi-one-dimensional wires (solid line) and in open quantum dots
with $\Nl=\Nr\gg 1$ (dashed line).  The maximum
in $\rm{rms}(B)$ occurs at $k_BT \sim E_c$ and vanishes at low ($k_BT \ll E_c$)
and high ($k_BT \gg E_c$) temperatures.
}
\label{fig:4}
\end{figure}

\section{Thermopower }
For completeness we make a few brief comments about the thermopower
coefficient, which
has been treated in Refs. \cite{AAAZ,LK88} for the case of a disordered
wire and in Ref. \cite{LSB}
for the case of a chaotic quantum dot.  The thermopower $Q = -B/G$ is
the coefficient of the voltage induced by a temperature difference when
no electrical current
flows, $I_{\rm e}=0$.  To leading order in the Sommerfeld expansion $Q=0$
and at first order $Q \sim (\kB T /e) G^{\prime}_\mu /G_\mu  \sim
(\kB T/\varef)$ for a macroscopic conductor.  For a mesoscopic conductor
$G^{\prime}_\mu /G_\mu  \sim (1/g\Ec )$ and the thermopower fluctuates
in sign and  for $\kB T \sim \Ec $ is  enhanced by a factor
$(\varef/\Ec )g^{-1}$ which is typically several orders of magnitude.
The fluctuations of $Q$ to the lowest order in $1/g$ are
determined by fluctuations of $B$, the latter can be
calculated from \req{Cdef} for $C_{11}$:
\be
\label{TP}
{\rm var} (B) = \left(\frac{\kB G_0}{e}\right)^2 C_{11}
\equiv \left(\frac{\kB G_0}{e}\right)^2 h^2\left(T, \Ec\right).
\ee

The behavior of ${\rm rms}( B )=(\kB G_0/e) h(T,\Ec)$ as
a function of $\kB T/\Ec $ is similar but not identical
to that of ${\rm var}(\eta)$, increasing
linearly with $T$ for $\kB T \ll \Ec $ and then decaying as
$T^{-1/2}$ for $\kB  T \gg \Ec $, see Fig.~\ref{fig:4}.
For diffusive wires $h(T,\Ec)$ was calculated in Ref.~\cite{LK88}.
Here we present the low and high
temperature asymptotes of $h(T,E_{\rm c})$:
\begin{subequations}
\begin{eqnarray}
h(T,E_{\rm c}) & = &  \frac{2}{3} \sqrt{10\zeta(8)}
\frac{\kB T }{\Ec }, \quad  \kB T\ll \Ec,\\
h(T,E_{\rm c}) & = &    \sqrt{\frac{\pi^2-6}{27\pi}}
\sqrt{\frac{\Ec}{\kB T}}, \quad  \kB T\gg \Ec
\end{eqnarray}
\end{subequations}
with $\Ec=D/L^2$.
The maximum  of $h(T,\Ec)$ is $h^{\rm max} \approx 0.17 $
at $\kB T\approx 0.37 \Ec $. For geometry of an open quantum dot,
$\Ec=\Nch\mls/(2\pi)$ and
the asymptotes are
\begin{subequations}
\begin{eqnarray}
\!\!\! \!\!\!\!\!\!
h_{\rm qd}(T,\Ec) \!\!\! & = &   \!\!\! \frac{\sqrt{2}\pi^2}{3}\frac{\Nl \Nr }{\Nch^2}
\frac{\kB T}{\Ec}, \quad  \kB T\ll\Ec,
\label{hqdlt}\\
\!\!\! \!\!\!\!\!\!
h_{\rm qd}(T,\Ec) \!\!\! & = &    \!\!\! \sqrt{\frac{\pi^3-6\pi}{18}}
\frac{\Nl \Nr }{\Nch^2}
\sqrt{\frac{\Ec}{\kB T}}, \  \kB T\gg\Ec.
\end{eqnarray}
\end{subequations}
The maximum of $h \approx 0.66 \Nl\Nr/\Nch^2$
occurs at $\kB T\approx 0.49 \Ec $. The low temperature
asymptote, \req{hqdlt},
was previously found in Ref.~\onlinecite{LSB}.

The maximum fluctuations in thermopower are
$Q_{max} \sim (\kB /e) g^{-1}$ at $\kB T \sim \Ec $.
The thermopower corrections to the thermal conductance $G_{\rm h}$
are always smaller than the direct quantum correction
by a factor $g^{-1}$ and therefore only play a significant role
in the violation of the WF law when $g$ approaches unity.

\section{Conclusions}

In conclusion, we evaluated the variance of the ratio of the heat
and electric conductances in mesoscopic systems. We showed that this
ratio
does not fluctuate as $T \to 0$ and exactly satisfies the
Wiedemann-Franz law, but
fluctuates with a finite
variance for non-zero $T$ and is a non-monotonic function of temperature with a
maximum at temperature equal to the Thouless energy.
The ratio of the $e/h$  harmonic of the heat and electric conductance
correlation
functions in an Aharonov-Bohm geometry satisfies the Weidemann-Franz
law only at low
temperatures; at high temperatures their ratio becomes $0.55 l_0^2$.

We thank Venkat Chandrasekhar for discussion of his recent
experiments. This work was supported by the W. M. Keck Foundation and
by NSF Materials Theory grant DMR-0408638.

%%%\bibliography{WFL}

\end{document}